\documentstyle[12pt,preprint,lscape]{aastex}

\begin{document}

\newcommand{\zdot}{\makebox[0pt][l]{.}}
\newcommand{\up}[1]{\ifmmode^{\rm #1}\else$^{\rm #1}$\fi}
\newcommand{\dn}[1]{\ifmmode_{\rm #1}\else$_{\rm #1}$\fi}
\newcommand{\upd}{\up{d}}
\newcommand{\uph}{\up{h}}
\newcommand{\upm}{\up{m}}
\newcommand{\ups}{\up{s}}
\newcommand{\arcd}{\ifmmode^{\circ}\else$^{\circ}$\fi}
\newcommand{\arcm}{\ifmmode{'}\else$'$\fi}
\newcommand{\arcs}{\ifmmode{''}\else$''$\fi}

\pagestyle{plain}

\def\thefootnote{\fnsymbol{footnote}}

\title{Difference Image Analysis of the OGLE-II bulge data. II.
Microlensing events.}

\author{P.R. Wo\'zniak$^{1,2}$, A. Udalski$^3$, M. Szyma\'nski$^3$, M. Kubiak$^3$,
G. Pietrzy\'nski$^{3,4}$, I. Soszy\'nski$^3$, K. \.Zebru\'n$^3$
}

\affil{$^1$ Princeton University Observatory, Princeton, NJ 08544, USA}
\affil{$^2$ Los Alamos National Laboratory, MS-D436, Los Alamos, NM 87545, USA}
\affil{e-mail: wozniak@lanl.gov}

\affil{$^3$ Warsaw University Observatory, Al. Ujazdowskie 4, 00-478 Warszawa, Poland} 
\affil{e-mail: (udalski,msz,mk,pietrzyn,soszynsk,zebrun)@astrouw.edu.pl}

\affil{$^4$ Universidad de Concepci\'on, Departamento de Fisica,
Casilla, 160-C, Concepci\'on, Chile}
\affil{e-mail: pietrzyn@hubble.cfm.udec.cl}

\begin{abstract}

We present a sample of microlensing events discovered in the Difference
Image Analysis (DIA) of the OGLE-II images collected during 3 observing
seasons, 1997--1999. 4424 light curves pass our criteria on the
presence of a brightening episode on top of a constant baseline.
Among those, 512 candidate microlensing events were selected visually.
We designed an automated procedure, which unambiguously selects up to
237 best events. Including 8 candidate events recovered by other
means, a total of 520 light curves are presented in this work.

In addition to microlensing events, the larger sample contains certain types
of transients, but is also strongly contaminated by artifacts. All
4424 light curves in the weakly filtered group are available
electronically, with the intent of showing the gray zone between
microlensing events and variable stars, as well as artifacts, to some
extent inevitable in massive data reductions. We welcome suggestions
for improving the selection process before the full analysis of
complete 4 seasons of the OGLE-II bulge data. Selection criteria for
binary events can also be investigated with our extended sample.

\end{abstract}

\section{Introduction}

\label{sec:intro}

Microlensing, and its potential to probe the mass distribution of compact
dark objects in the Galaxy, were first proposed as an observational
subject in mid eighties (Paczy\'nski 1986). Since the first discoveries in 1993
Galactic microlensing has grown into a whole new field of astrophysics
with several teams conducting large monitoring programs of the densest stellar
fields of the sky (see Paczy\'nski 1996 for a review).
The cross section to detectable light magnification by a stellar mass
lens in the Galaxy is of the order $10^{-6}$ in case of typical sources
in the Milky Way bulge, and $10^{-7}$ for sources in LMC and SMC. If we
want to maintain practical event rates, this leads to monitoring millions
of stars over extended time intervals. The main goal of these efforts still
remains unchanged: to provide large samples of microlensing events with
the best possible understanding of involved systematics. The line of sight
towards the Galactic bar returns most events, and several estimates of the
optical depth based on conventional PSF photometry are already available.
Early estimates came from OGLE-I (9~events, Udalski et al. 1994a) and MACHO
(45 events, Alcock et al. 1997), and both gave a somewhat unexpectedly large
value $\sim3.5\times10^{-6}$ with 30--50\% uncertainty.
Popowski et al. (2000) emphasized that the available information is
insufficient for a robust comparison with models. Only continued effort can
improve this situation. Their optical depth of $(2.0\pm0.5)\times10^{-6}$,
based on 52 events with red clump giants as sources, should be less prone
to blending and partially alleviates the problem.

Measuring the mass spectrum of lenses will most likely require individual
mass estimates using interferometric follow up (possibly from space)
of the most promising events, for which the degeneracy between physical
lens parameters can be broken (Gould 1995 and references therein).
Nevertheless, with photometry alone for about 1500 well covered events
with good signal to noise and well assessed blending, one could
normalize model components of the inner galaxy by comparing
corresponding maps of predicted optical depth with the observed number
of events across the Galactic bar (Han \& Gould 1995).

The Optical Gravitational Lensing Experiment, after its upgrade to OGLE-II
in 1997, collected data with the new, dedicated 1.3 m Warsaw Telescope
at the Las Campanas Observatory. OGLE-II database contains 4 seasons of
data, a total of 15,500 $I$ band images in 49 bulge fields covering
11 square degrees. The Early Warning System (Udalski et al. 1994b), adapted
to OGLE-II, reached the rate of 78 events per year discovered in real time
during the observing season of 2000 and posted on the OGLE web page at
{\it http://www.astrouw.edu.pl/$\sim$ftp/ogle/ogle2/ews/ews.html}.
A catalog of 214 events from the off line search of 3 years of data was
published by Udalski et al. (2000). Presently OGLE has upgraded its
2k$\times$2k pixel camera to a mosaic of 8 2k$\times$4k CCDs. By scaling
the number of detections, we can expect up to 1000 events per year from
OGLE-III, starting in mid 2001.

Limitations associated with photometry in extremely crowded fields stimulated
the development of difference imaging techniques, specifically their
adaptations for microlensing. Currently, two algorithms have successful large
applications: Fourier division (Tomaney \& Crotts 1996, Alcock et al. 1999a)
and linear kernel decomposition in real space (Alard \& Lupton 1998,
Alard 2000, Wo\'zniak 2000, and more recently Bond et al. 2001).
Difference Image Analysis (DIA) offers numerous advantages over
conventional PSF photometry for variable objects in crowded environments.
After matching PSFs of two images of the same stellar field, the difference
flux corresponding to these images is no longer crowded. There is no need
for fitting multi-PSF models, which is usually a degenerate procedure
and results in strongly non-Gaussian error distributions. In fields as
crowded as the Galactic bulge, typically several stars contribute light to a single
observed PSF. DIA removes biases in the light centroid of the variable
star. It also provides generally much better S/N ratio, which results in more
detections. Alcock et al. (2000b) obtained a supposedly less biased value of
the optical depth in Baade's Window using DIA: $(3.2\pm0.5)\times10^{-6}$,
in better than just formal agreement with the early work. However,
they analyzed only 8 out of 94 MACHO fields.

In this paper we present a sample of microlensing events discovered
in a complete reanalysis of 3 seasons of the OGLE-II bulge data using
DIA photometry based on the algorithm of Alard \& Lupton (1998) and
Alard (2000). Our main purpose is to eliminate human
judgment, and design a fully automatic selection procedure, which has not
been accomplished in OGLE-II so far (Udalski et al. 2000). In Section~\ref{sec:search}
we present a programmable algorithm, which unambiguously selects events best
interpreted as microlensing upon visual inspection. This step is essential
in determination of the efficiency function to detection of microlensing
events, a necessary ingredient of the optical depth estimates. Additionally,
we discuss a possible source of contamination of the OGLE-II and other
microlensing samples, the existence of artifact light curves. With the
released sample of light curves, the confusion between microlensing,
variability of other kinds and instrumental artifacts can be studied.
Such training data can also be used to refine classifiers of single
microlensing events and, even more interestingly, to attack the problem
of automated selection of binary and other exotic events.

\section{Data}

\label{sec:data}

All OGLE-II frames were collected with the 1.3 m Warsaw Telescope
at the Las Campanas Observatory, Chile. The observatory is operated by the
Carnegie Institution of Washington. The ``first generation'' OGLE camera uses
a SITe~3~~ $2048 \times 2048$ CCD detector with $24 \mu$m pixels resulting
in 0.417$\arcs$ pixel$^{-1}$ scale. Images of the Galactic bulge are taken in
driftscan mode at ``medium'' readout speed with the gain 7.1 $e^-$/ADU
and readout noise of 6.3 $e^-$. Saturation level is about 55,000 ADU.
For the details of the instrumental setup, we refer the reader to
Udalski, Kubiak \& Szyma\'nski (1997).

\begin{figure}[Ht]
\vspace{14cm}
\includegraphics{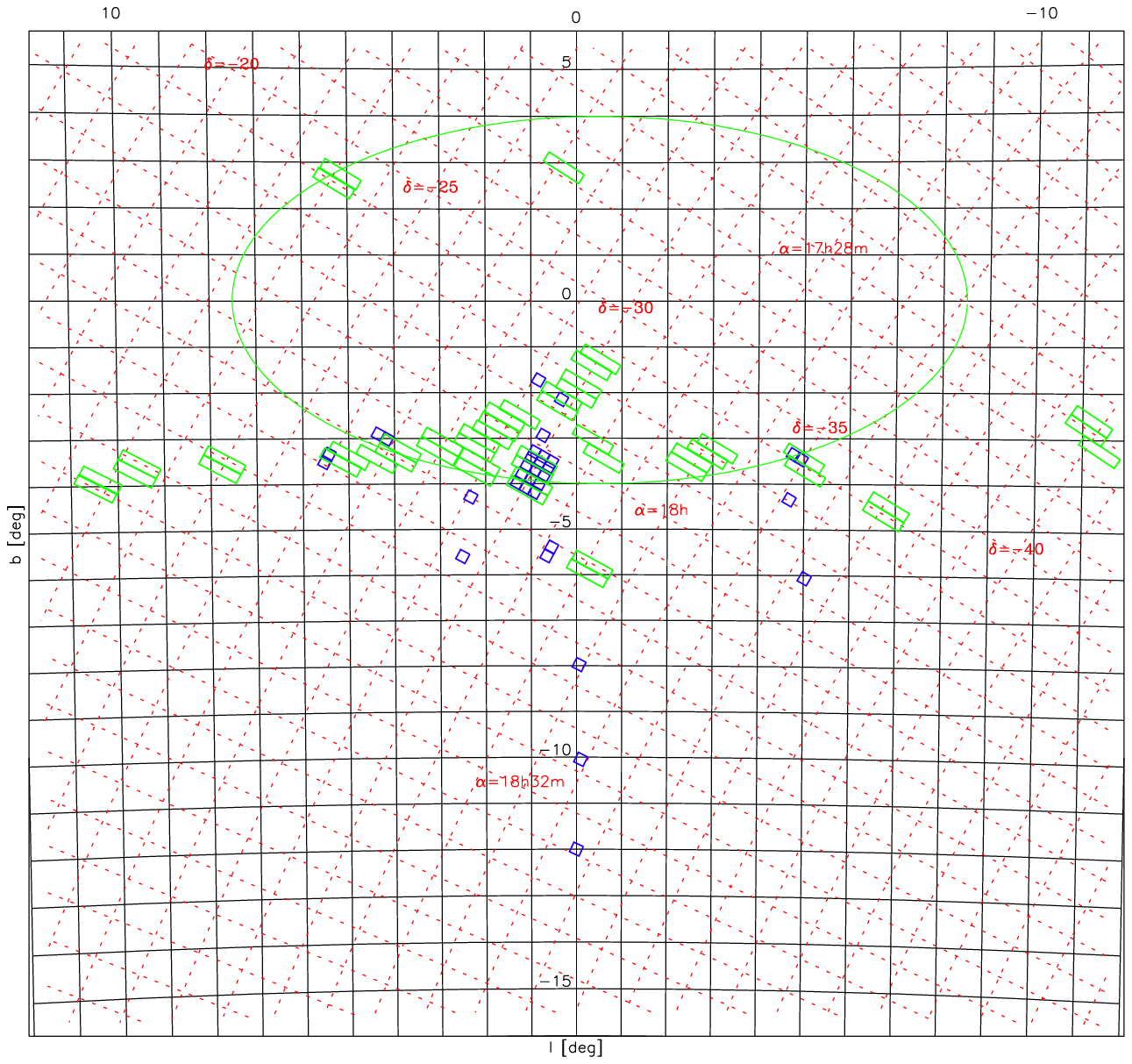}
\caption{\label{fig:fields}
OGLE bulge fields in galactic coordinates (gnomonic projection, great
circles are mapped to straight lines).
Green strips are the OGLE-II scans and blue squares are the old OGLE-I fields.
Large oval indicates the location of the Galactic bar. Fields are selected
in windows of low extinction and avoid very bright foreground stars.
}
\end{figure}

The majority of frames was taken in the $I$ photometric band. Effective
exposure time is 87 seconds. During observing
seasons of 1997--1999 the OGLE experiment collected typically between
200 and 300 $I$-band frames for each of the 49 bulge fields BUL\_SC1--49.
OGLE-II images are 2k$\times$8k strips, corresponding to
$14\arcm\times 57\arcm$ in the sky, therefore the total covered area of
the bulge is about 11 square degrees. The number of frames in $V$ and $B$
bands is small and we do not analyze them with the DIA method. The median
seeing is 1.3$\arcs$ for our data set. In Table~\ref{tab:fields} we provide
equatorial and galactic coordinates of the field centers and the total number
of analyzed frames. Figure~\ref{fig:fields} schematically shows
locations of the OGLE-II
bulge fields with respect to the Galactic bar. Fields BUL\_SC45 and BUL\_SC46
were observed much less frequently, mostly with the purpose of maintaining
phases of variable stars discovered by OGLE-I. Observations of fields
BUL\_SC47--49 started in 1998; the season of 1997 is not available for them.

\section{DIA photometry of OGLE-II bulge frames}

\label{sec:phot}

The DIA data pipeline we used is based on the image subtraction algorithm
developed by Alard \& Lupton (1998) and Alard (2000), and was written by
Wo\'zniak (2000). Processing of large 2k$\times$8k pixel frames is done
after division into 512$\times$128 pixel subframes with 14 pixel margin
to ensure smooth transitions between coordinate transformations and fits
to spatially variable PSF for individual pieces. Small subframe size allows
us to use polynomial fits for drift scan images, in which PSF shape and local
coordinate system vary on scales of 100--200 pixels. The reference image,
subtracted from all images of any given field, is a stack of 20 best images
in the sequence.

We adopted kernel expansion used by Wo\'zniak (2000), generally applicable
to all OGLE-II data. The kernel model, represented by a 15$\times$15
pixel raster, consists of 3 Gaussians with sigmas 0.78, 1.35, and 2.34 pixels,
multiplied by polynomials of orders 4, 3, and 2 respectively. The pipeline
delivers a list of candidate variable objects and their difference light
curves. The initial filtering is very weak, with only minimum of assumptions
about the variability type. Candidate variables are flagged as ``transient''
or ``continuous'' variables depending on whether variability is confined to
episodes in an otherwise quiet object, or spread throughout the observed time
interval. The total number of candidate variables in all 49 fields was slightly
over 220,000, including 150,000 ``continuous'' and 66,000 ``transient''
cases.
Only 4600 objects passed both filters, confirming sensible definitions
of classes. This initial classification is refined in
Section~\ref{sec:transients}. Number of detected variable objects in a given
field depends on the number density of stars, extinction, and number of
available measurements. It ranged from about 800 to over 9000 per field.
Variability database for all 49 fields comprises 2 GB of data.

The error distribution in measurements from our DIA pipeline is nearly
Gaussian with the average scatter only 17\% above the Poisson limit for faint
stars near $I$=17--19 mag, gradually increasing for brighter stars, and reaching
2.5 times photon noise at $I\sim11$ mag (about 0.5\% of the total
flux). Error bars adopted in this paper are photon noise estimates
renormalized using the curve of Wo\'zniak (2000). Because the method
effectively monitors all pixels for possible variability, microlensing
events may be discovered even where no object is detected in the
reference image. In regular searches monitoring is conducted only
for objects detected in a single good quality image, a template.
This issue is related to centroid finding. Currently in our DIA
pipeline centroids are calculated based on the variable signal in a number
of frames. As a result the centroid will be poorly known for an object
with low S/N variability, even if it is very bright on the reference image.
Ideally one would use both pieces of information. It is usually obvious how
to determine the centroid in the presence of blending when confronted
with one particular object of interest, but an optimal algorithm for
extracting all variability in the field using DIA is yet to be developed.

\section{Search for events}

\label{sec:search}

Microlensing event selection algorithms of leading surveys are very involved,
multi-step procedures with numerous vaguely justified fudge factors.
The underlying problem is that the error distributions in photometry are
usually not Gaussian and poorly controlled. In the MACHO project several
levels of cuts aim solely at filtering known types of variable stars,
and the $\chi^2$ statistic based on formal error bars is weighted with
the event amplitude (see Alcock et al. 2000b for criteria with the DIA
photometry). EROS team came up with the algorithm, in which short
period variables are rejected based on large deviations of points from
linear interpolation between the two neighboring measurements
(e.g. Afonso et al. 1999). Both teams make use of color information and
approximate achromaticity of microlensing. Simultaneous observations
with more than one filter provide a strong veto against artifacts.
Dwarf novae are sufficiently blue during outbursts to be rejected
based on their colors (Alcock et al. 1999b). Unfortunately
the number of the OGLE-II observations in $V$ and $B$ bands is very small
compared to $I$, and cuts based on achromaticity are out of the question.

Wo\'zniak (2000) made a preliminary statement that using the DIA method
512 microlensing events were found in 3 observing seasons of data for
all 49 OGLE bulge fields. 305 of those were recovered in an
algorithmic search. At the time, however, the detailed noise
properties of the sample were not known and the old algorithm
was fairly artificial. In particular, the properties of the Gaussian
distribution of
errors were not explicitly used. We abandon this approach and develop
an entirely new set of criteria with less emphasis on the number of
events, more attention to reducing the number of involved parameters,
and better suited for approximately Gaussian noise in our DIA photometry.

In the early days of OGLE-I A. Udalski was the first to fully appreciate
that one of the important variability characteristics of microlensing events
is that most of the time there is no variability at all. The main selection
criterion for microlensing in OGLE requires that the star is formally constant
during at least one observing season and varies only within a localized
time interval (Udalski et al. 1994a).
Despite the fact that this cut shortens the list of
candidate light curves by a large factor, so far the final selection of events
in OGLE-II still involved visual inspection of many candidates
(Udalski et al. 2000). In this work we attempted to define selection cuts
which grasp the main idea of a long constant baseline, but also
exploit good noise properties of the DIA measurements allowing a better
use of $\chi^2$ related statistics.

\subsection{Transients}

\label{sec:transients}

Before the main selection process, light curves are cleaned of suspicious
points. Occasional problems with bleeding columns and undetected cosmic rays
produce points with irrelevant values, which can be spotted using several
parameters supplied by the pipeline. First, we remove points which came
from poor difference frames with $\chi^2$ per pixel above 6.0. Measurements
with 4 or more saturated pixels within the fitting radius are also removed.
Next, we determine the median photometric error of all remaining measurements
and median $\chi^2$ per pixel value of the corresponding PSF fits. We reject
the points for which either the photometric error or $\chi^2$ per pixel
is a factor of 10 larger than the median. Usually almost all measurements
are admitted, nevertheless these filters are necessary to prevent
loss of good events due to a few bad points. Finally, we reject stars with
centroids determined using fewer than 4 frames and the total number of
useful measurements below 50\% of all frames for a given field. This removes
numerous stars very close to one of the frame edges.

In the next step, we check for the presence of a constant baseline in the light
curve. We define a running window of length equal to 50\% of all usable
measurements, and select the position of the window which minimizes
$\sigma_{F_{\rm base}}$, the scatter around the mean $F_{\rm base}$. In order
to avoid discrimination against both ends of the observed period,
it is assumed that after the last point the light curve continues starting
over from the beginning. To check for possible variability in the baseline
we bin the window into 10 point intervals and calculate the mean $F$ and
$\sigma_F$ in each bin separately. We require that all but maximum 2
individual means stay within $\sigma_{F_{\rm base}}$ of $F_{\rm base}$.
Also the mean of individual $\sigma_F$ values in all bins cannot be lower
than $0.5\sigma_{F_{\rm base}}$, to exclude smooth, long term
variability within the baseline.

We also define a dynamic threshold for interesting brightening episodes.
If $N_{0-1\sigma}$ is the number of points within $1\sigma_{F_{\rm base}}$
of $F_{\rm base}$, and $N_{1-3\sigma}$ the number of points between 1 and 3
$\sigma_{F_{\rm base}}$ of $F_{\rm base}$, the expected ratio
$N_{1-3\sigma}/N_{0-1\sigma}$ is about 32/68 for a Gaussian distribution.
Larger ratios suggest the presence of non-Gaussian wings, which may result
from intrinsic variability of the source. If $\sigma_{\rm ph}$
is the mean photometric error estimate for points in the baseline window,
the ratio ${\sigma_{F_{\rm base}}/\sigma_{\rm ph}}$ should be near 1 in the
absence of variability. Intrinsic variability or atypically crowded
environment could result in larger values. At this stage we admit low
amplitude variables with brightening episodes and leave the margin for
some scatter around the average baseline properties by increasing
the threshold.

\noindent We adopt threshold $\delta$:

\begin{eqnarray}
\delta&=&\sigma_{F_{\rm base}}\times\delta_1\times\delta_2 , 
\end{eqnarray}

\noindent
where

\begin{eqnarray}
\delta_1&=&\max\left\{ {N_{1-3\sigma}/N_{0-1\sigma} \over 32/68},
1 \right\}  \\
\delta_2&=&\max\left\{   {{\sigma_{F_{\rm base}} \over \sigma_{\rm ph}}},
1 \right\} 
\end{eqnarray}

\noindent
The limit $\sigma_{F_{\rm base}}/\sigma_{\rm ph} < 2.5$ is set on the allowed
scatter in the baseline for all objects flagged initially by the pipeline as
``transient'' type (Section~\ref{sec:phot} and Wo\'zniak (2000)).
Candidate variables flagged as ``continuous'' do not enter the main
study and are treated separately (Section~\ref{sec:events}).

Once the presence of a constant baseline is established, we look for
brightening episodes in the light curve using threshold $\delta$. We declare
that there is an episode if N consecutive points deviate by ${\rm n}\delta$
up or down with respect to $F_{\rm base}$. We consider two types of episodes
with (N, n$\delta$) values of (3, 5$\delta$) and (4, 4$\delta$). It is
required that the candidate transient event has 1 or 2 separate episodes
above the baseline flux within at least one of the above types. No episodes
below the baseline flux are allowed. The last requirement removes some
interesting variables, which may be a subject of a separate study.

A total of 4424 light curves went through these cuts, roughly 2\% of the
group initially suspected of variability. Most of them have episodes with very
few points at low signal to noise, and therefore very poorly known centroids.
Another numerous type which escapes filtering is very slow and low signal to noise
trend. Some of those could be due to proper motions just at the limit of
detection (Eyer \& Wo\'zniak 2000). Instrumental artifacts are a serious problem
in all microlensing searches due to the amount of data which has to be
filtered for each discovered event. This topic is usually ignored in papers.
We find that real stellar variability background is easier to separate from
microlensing than artifacts, which account for most of our 4424
``transients''. A group of candidate variables with close positions
on the frame and similar looking light curves is a clear indication
of a problem. Unfortunately the wings of very bright, saturated stars
can generate ghost variables within many tenths of pixels due to changing
weather conditions and their own variability. Nevertheless, the final
list of ``transients'' contains many interesting objects, including microlensing
events. We conclude that prior knowledge about the type of variability
is essential for effective searches. In particular, microlensing
model fits are critical for an automated detection of events in the presence
of contaminating background of artifacts.

\subsection{Microlensing events}

\label{sec:events}

To make sure that we include binary lensing events and other interesting
departures from classic single point mass microlensing, a visual search
through light curves from the previous cut is performed first. We made
fits of the single microlensing curve for all 4424 candidates from
Section~\ref{sec:transients}. The model in the form:

\begin{eqnarray}
M(t)&=&F_0 + F_1\times(A(t)-1)  ,
\end{eqnarray}

\noindent
where

\begin{eqnarray}
A(t)&=&{{u^2+2}\over u~\root\of{u^2+4}}  \\
\nonumber \\
u^2&=&u_0^2 + {\left({t-t_{\rm max}\over t_0}\right)}^2  ,
\end{eqnarray}

\noindent
includes the unknown zero point $F_0$ in addition to the lensed flux $F_1$,
the impact parameter $u_0$, the Einstein ring crossing time $t_0$, and the moment
of maximum $t_{\rm max}$. By visual comparison with the fit, we recovered
512 light curves, for which the best interpretation is microlensing.

Then, we defined several parameters of the fit, and in the so constructed
parameter space we examined the locus of our visually selected sample
compared to the remaining episodic light curves. Our efforts to
quantify what makes observers believe that they are seeing microlensing
resulted in the following conclusion: in addition to a well defined
baseline, the next important property is the presence of several consecutive
high S/N points which are well fitted with the microlensing curve.
Therefore, the main parameter distinguishing single microlensing events
from other types of variability in our search is:

\begin{eqnarray}
R&=&\max\limits_k
{\left[ {\sum\limits_{\scriptscriptstyle i=k}^{\scriptscriptstyle k+11}
        (f_i - F_0)^2} \over
        {\sum\limits_{\scriptscriptstyle i=k}^{\scriptscriptstyle k+11}
        (f_i - M_i)^2} \right]}^{1/2}  ,
\end{eqnarray}

\noindent
where $f_i$ is the flux of the $i$-th observation, $F_0$ is the fitted
baseline flux, $M_i$ is the best fit
microlensing model, the sums are over a 12 point window, and $R$ is the
maximum value of the ratio over the entire light curve. In case of a good
fit $R$ is roughly proportional to S/N ratio, while for a bad fit
$R$ is much lower, even in cases of high S/N variability. Again, we
formally run the test on cyclic light curves with the beginning and the end
joined together.

As emphasized before, in microlensing light curve one expects a baseline which
is long compared to the event duration. The use of total $\chi^2$ or total
$\chi^2$/d.o.f. is not very efficient in this case. It is important to have a
handle on the goodness of fit separately near the event and far from the event,
in the baseline. We define the event region of the light curve, where the flux
from the best fit model is at least 1 median error $d_{1/2}$ above
$F_0$. We calculate two approximately $\chi^2$ distributed sums:

\begin{eqnarray}
s_1&=&{1 \over {(m_1 - 4)}}
\sum\limits_{\scriptscriptstyle M_i-F_0>d_{1/2}}
{(f_i - M_i)^2 \over \sigma_i^2 } 
\end{eqnarray}

\noindent
for $m_1$ points in the event region and

\begin{eqnarray}
s_2&=&{1 \over {(m_2 - 1)}}
\sum\limits_{\scriptscriptstyle M_i-F_0<d_{1/2}}
{(f_i - M_i)^2 \over \sigma_i^2 } 
\end{eqnarray}

\noindent
for $m_2$ points of the baseline region, both of which should stay near
1 for normal cases of microlensing and properly normalized errors.
We also consider total duration of the event $T$, the time between the
beginning and the end of the event region, as given by the best fit model.
Long lasting events are ambiguous, because their baselines are short relative
to the duration of the experiment. This notion can be quantified by
setting a higher S/N threshold for longer events.

\begin{figure}[t]
\vspace{12cm}
\includegraphics{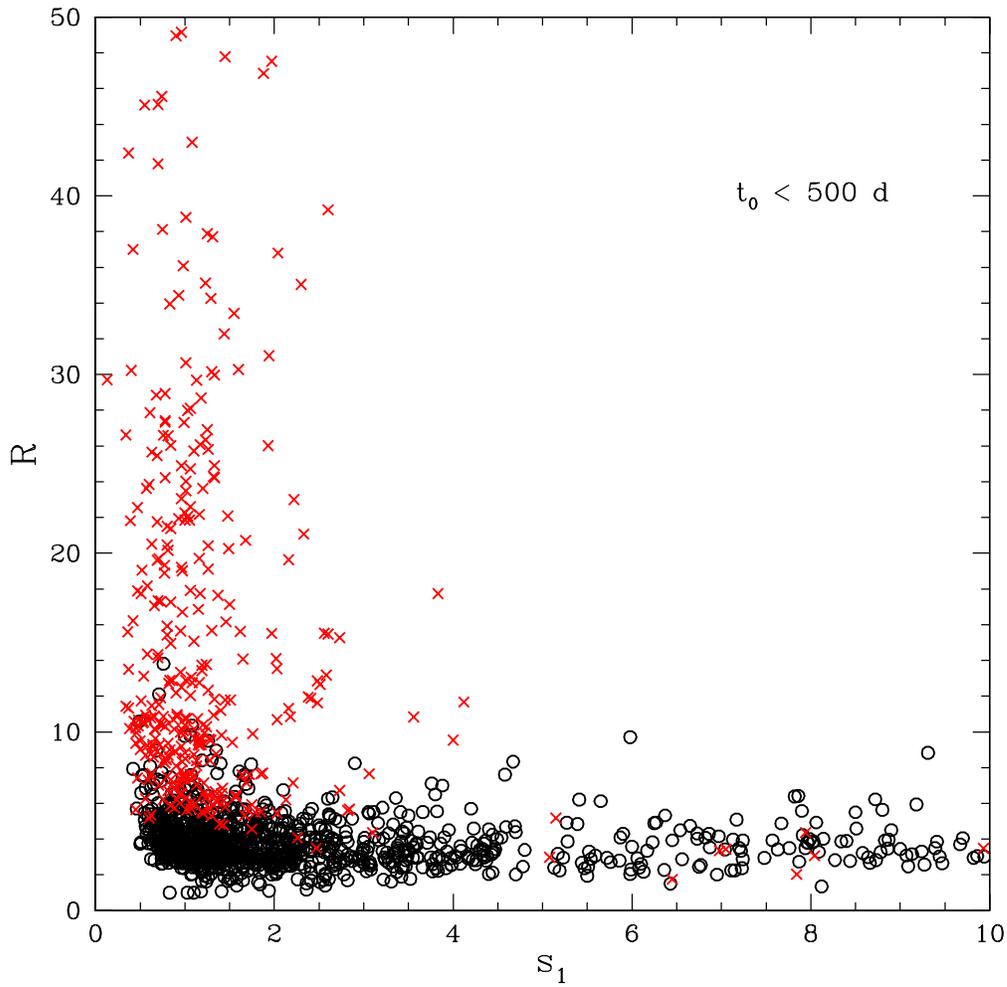}
\caption{\label{fig:selection}
Distribution in the $(s_1, R)$ plane of microlensing events (red crosses)
visually selected among the light curves which passed the automated selection
criteria for ``transients'' (black circles). DIA photometry allowed a great
reduction of the region where microlensing events can be easily confused with
other effects.
}
\end{figure}

\begin{figure}[t]
\vspace{12cm}
\includegraphics{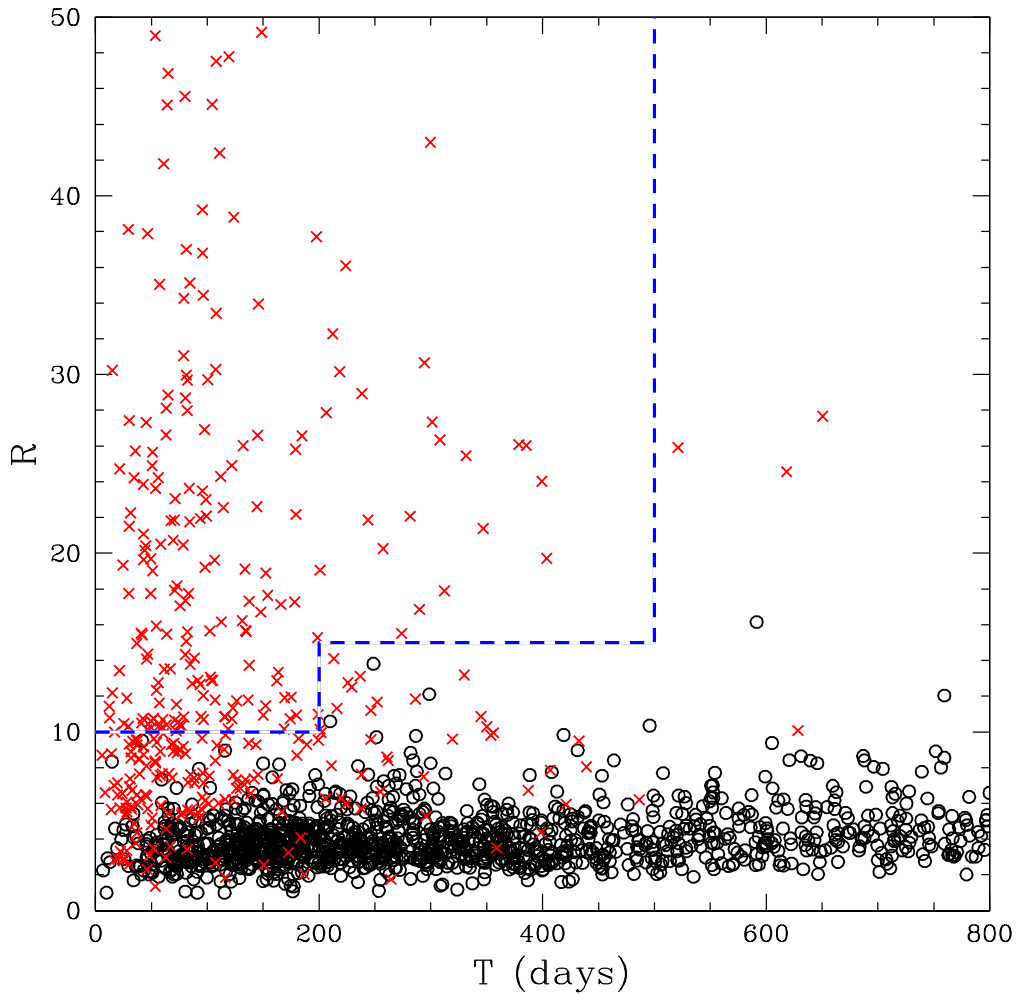}
\caption{\label{fig:duration}
Distribution of visually selected microlensing events (red crosses)
and automatically selected ``transient'' light curves (black circles)
in the $(T, R)$ plane. Selection based on $R$ alone (S/N ratio
and goodness of fit) becomes more uncertain for events with long
durations $T$. The blue dashed line corresponds to the final selection cut
adopted in Section~\ref{sec:events}.
}
\end{figure}

In Figure~\ref{fig:selection} we show the distribution in the $(s_1, R)$
plane of candidate light curves with $s_2<2.0$, and $0<T<500$ days,
and minimum 8 high points in the event region. A high point is simply
the one detected at more than $3\sigma$ above the baseline. Single
microlensing events are relatively well separated by $R$ alone, with all
remaining types of variability and artifacts located at low $R$ and spanning
a large range in $s_1$. Above $R=15$ there are no light curves of unwanted types,
and there are 134 events with reasonably good fits, that is $s_1<2.0$.
It is possible to decrease the limit on signal to noise to $R=10$ for shorter
events with $T<200$ days, as shown in Figure~\ref{fig:duration}.
This relaxed cut selects 188 events.

The above criteria strongly reject events with any departures from a single
point mass microlensing model. In Figures~\ref{fig:selection} and
\ref{fig:duration} some events from the visual sample are located
in the region of $R<10$ and have poor fits with $s_1>3.0$.
These are mostly cases of binary lensing and possibly binary sources. Other
departures from a perfect fit to the model are expected too. The cumulative
error distribution in Wo\'zniak (2000) revealed a weak non-Gaussian tail
at the level of $\sim1$\%. Outlying points can also be produced by
parallactic effects and weakly binary events with very low mass companions,
like Jovian planets. Examples of possible planetary disturbances
include: SC20 1793, SC29 2054, SC36 7980, and SC39 3160. In the
present sample their nature can only be studied statistically. A direct
detection will require much more frequent observations, which are
planned for some of the fields in OGLE-III.
Such events should be included in the determination
of the optical depth. Therefore, we perform iterative procedure, in which
we reject the worst point in the fit, obtain a new fit and reexamine
all parameters of the event. There are 237 events which pass the above
defined criteria after between 0 and 5 consecutively worst points are removed.
All light curves selected in this way were also selected visually. We
show these best events in Figure~\ref{fig:auto_lens}. Figure~\ref{fig:vis_lens}
shows light curves of the remaining candidate events selected visually
from the group which passed the criteria in Section~\ref{sec:transients}.
For those, the uncertainty of classification is generally larger.
The corresponding model fits and some basic data are presented in
Tables~\ref{tab:auto_lens} and \ref{tab:vis_lens}. Columns are:
1. OGLE field number, 2. star number in the DIA variability database,
3. star number in the OGLE database for events cross identified with
the OGLE catalog of events, 4. 5. equatorial coordinates, 6. base line
difference flux, 7. lensed flux, 8. impact parameter, 9. Einstein ring
crossing time, 10. moment of maximum light (JD Hel.$-$2450000),
11. base line $I$ magnitude, 12. $f$, the ratio of the lensed flux to the total
base line flux. Columns 6.--10. are best fit model parameters. Columns
11.--12. are derived from 6.--7. using calibration data distributed
with light curves (Section~\ref{sec:discussion}). The fits are preliminary
in a sense that the near degeneracy of the microlensing curve
occasionally results in unphysical values of the best fit parameters.
In such cases a careful study using additional information is usually
needed in order to find a better model. 

We correlated positions of the candidate variables in our DIA data with
the positions of 214 microlensing events reported by Udalski et al. (2000).
The nearest candidate within a 2 pixel radius (0.8$\arcs$) was adopted
as a positive identification. 127 events out of 237 in our algorithmic
sample were identified, 39 more positive matches came out of the visual
sample. In the process we recovered 8 additional
events from the DIA variability data (Table~\ref{tab:other_lens}
and Figure~\ref{fig:other_lens}). Three of them actually passed the
criteria for ``transients'' (Section~\ref{sec:transients}) but have not been
picked up visually: BUL\_SC3 7783, SC5 4321 and SC37 8176. Three more
events failed the test for a constant base line because of their extremely
long duration, and two remaining ones had the base line scatter above
the limits. SC44 1980 seems to be a genuine variable which went
through a microlensing event. OGLE database identifications for 174
events are included in Tables~\ref{tab:auto_lens}, \ref{tab:vis_lens}
and \ref{tab:other_lens}. This leaves us with 40 events reported by
Udalski et al. (2000) which did not make it to the initial list of
suspected variables. Most of these light curves have peaks with very
low S/N ratio, however 6 events in this group were well measured
by the standard OGLE photometric pipeline. One of the selection
cuts for candidate variables in the DIA pipeline is based on how well
the PSF fits the residual in the so called variability image
(Wo\'zniak 2000). Most likely, the parameters of this cut need some
adjustment. As mentioned before, the optimal algorithm for extracting
variability from difference images is still a subject of research.

Selection of binary microlensing events has never been automated.
Some of the binary events pass the algorithmic criteria adopted
by the MACHO team, but the presence of a strong second peak is used by
MACHO to reject events (Alcock et al. 2000b). Binary fraction among
all events from the MACHO Alert System (Alcock et al. 2000a)
and visually selected samples from OGLE (Udalski et al. 2000), roughly
agree with what is known about binary stars (Mao \& Paczy\'nski 1991).
However, there are unexpectedly few reports of weak binaries without
caustic crossings compared to predictions of Di Stefano (2000),
suggesting a strong bias towards low impact parameters and a connection
to blending (Alcock et al. 2000a). In particular, our sample does not
include a single light curve, which consists of two well separated
point mass microlensing peaks. Such events should occur due to lensing
by wide binary systems, in which the separation between the two components
is several times larger than both Einstein ring radii.
Our sample of 4424 light curves provides some testing ground for a
more systematic study.

\section{Discussion and future work}

\label{sec:discussion}

We demonstrated that the process of selecting microlensing events in
OGLE-II data can be automated with the use of the DIA photometry. 
In summary, the main steps of the automated selection of microlensing
events are: identification of the constant baseline, detecting
an event region in the light curve with several high signal to noise
points, and finally, adjusting the threshold on S/N for the longest
(and therefore more ambiguous) events. Elimination
of human judgment in this procedure will allow a reliable determination
of detection efficiencies, and ultimately optical depth to microlensing.
In addition to the data for microlensing events found in this analysis,
the full set of light curves which passed much weaker constraints on
variability type (Section~\ref{sec:transients}), is available electronically
at {\it http://www.astro.princeton.edu/$\sim$wozniak/dia/lens}.
README file provides full details on light curves, calibration data
and other information which could not fit in this paper.
We encourage individuals to work with our weakly filtered data on
better selection criteria for single and binary microlensing events.
We also believe that it is important to show the gray zone
of microlensing surveys, where the background of variable stars, and even
more problematic background of artifacts, has to be separated from
the actual events.

Extremely large volumes of data processed by microlensing
surveys are very likely to hide interesting surprises. In case of the DIA
analysis of the OGLE-II BUL\_SC1 field, the strategy of releasing samples
selected using relaxed criteria has already paid off with the photometric
discovery of proper motions in very dense stellar fields
(Eyer \& Wo\'zniak 2000). Presently released set of microlensing
events has a much better controlled error distribution than the standard
DoPhot output, which makes it suitable for setting an upper limit
on the number of Jupiter mass planets around lensing stars.

The set of the OGLE-II images is closed. In the course of this project, the
last, fourth season of data was collected. All available images
will be reprocessed in the near future using pipeline of Wo\'zniak (2000).
We plan to adopt stricter approach to badly saturated stars by rejecting
larger areas around them. The cost of up to 10\% loss of the survey area
is still acceptable, as the intermediate samples of variable objects
will be much less contaminated by spurious variability.
In the mean time, we expect the final calibration of the OGLE-II database
of standard PSF photometry from the DoPhot based pipeline to be completed.
DIA light curves will be converted to standard Johnson--Cousins filters.
Difference imaging technique is being considered as the primary mode
of data reduction in OGLE-III, which started collecting
new images in mid 2001 using a mosaic of 8 2k$\times$4k CCDs. Still frame
images of OGLE-III will undoubtedly have much weaker spatial PSF variations
and should be easier to handle with the DIA photometry.

\acknowledgements

We would like to thank Prof. Paczy\'nski for constant support and
numerous insightful discussions. This work was supported
by the NSF grant AST-9820314 to Bohdan Paczy\'nski, Polish
KBN grant 2P03D00814 to A. Udalski, and 2P03D00916 to M. Szyma\'nski.
Additional support was provided by the Laboratory Directed Research
\& Development funds (X1EM and XARF programs at LANL).


\clearpage

\clearpage

\clearpage

\clearpage

\begin{deluxetable}{lccrrr}
\tablecaption{\label{tab:fields}OGLE-II bulge fields.}
\tablewidth{11.5cm}
\tablehead{
\colhead{Field} &
\colhead{$\alpha_{2000}$} &
\colhead{$\delta_{2000}$} &
\colhead{$l$} &
\colhead{$b$} &
\colhead{$N_{\rm obs}$} \\
&
\colhead{$\uph~~\upm~~\ups$} &
\colhead{$\arcd~~\arcm~~\arcs$} &
\colhead{$\arcd$} &
\colhead{$\arcd$}
}
\startdata
 BUL\_SC1 & 18:02:32.5 & $-$29:57:41 &    1.08 &  $-$3.62 &  197 \\
 BUL\_SC2 & 18:04:28.6 & $-$28:52:35 &    2.23 &  $-$3.46 &  192 \\
 BUL\_SC3 & 17:53:34.4 & $-$29:57:56 &    0.11 &  $-$1.93 &  309 \\
 BUL\_SC4 & 17:54:35.7 & $-$29:43:41 &    0.43 &  $-$2.01 &  324 \\
 BUL\_SC5 & 17:50:21.7 & $-$29:56:49 & $-$0.23 &  $-$1.33 &  307 \\
 BUL\_SC6 & 18:08:03.7 & $-$32:07:48 & $-$0.25 &  $-$5.70 &  228 \\
 BUL\_SC7 & 18:09:05.5 & $-$32:07:10 & $-$0.14 &  $-$5.91 &  219 \\
 BUL\_SC8 & 18:23:06.2 & $-$21:47:53 &   10.48 &  $-$3.78 &  211 \\
\enddata
\end{deluxetable}

\clearpage

\tablecaption{\label{tab:auto_lens}Microlensing events found in the
algorithmic search. See Section~\ref{sec:events} for details.
}

\clearpage

{\scriptsize

\begin{deluxetable}{lrrccrrrrrrr}
\tablewidth{17.5cm}
\tablehead{
\colhead{Field} &
\multicolumn{2}{c}{Star} &
\colhead{~~~~~~$\alpha_{2000}$~~~~~} &
\colhead{$\delta_{2000}$} &
\colhead{~~~~~~~$F_0$} &
\colhead{$F_1$} &
\colhead{$u_0$} &
\colhead{$t_0$} &
\colhead{$t_{\rm max}$} &
\colhead{~~~~~~~$I_0$~~} &
\colhead{~~~$f$~~~} \\
& \colhead{DIA} & \colhead{OGLE} & & & & & & & & &
}
\startdata
 sc1 & 1943 &        & 18:02:11.82 & -30:00:58.2 &   -16.8 &    85.0 & 0.3437 &  24.55 & 1323.47 & 17.130 &    0.287 \\
 sc1 & 2186 &        & 18:02:44.87 & -29:58:18.2 &     0.7 &    38.3 & 0.3650 &  17.25 &  989.99 & 19.078 &    0.769 \\
 sc2 &  854 &  27414 & 18:04:04.92 & -29:11:38.4 &   -16.1 &   208.1 & 0.5183 &   8.22 &  562.32 & 18.145 &    1.724 \\
 sc2 & 1799 &  65831 & 18:04:06.84 & -29:01:17.1 &  -104.2 &   601.2 & 0.3559 &  58.34 & 1457.29 & 14.580 &    0.197 \\
 sc2 & 2017 & 495787 & 18:04:42.37 & -28:59:38.8 &    -0.3 &    20.8 & 0.1338 &  38.52 & 1082.80 & 18.504 &    0.235 \\
 sc2 & 3095 &        & 18:03:58.77 & -28:47:39.3 &   -24.4 &    18.6 & 0.0214 &  43.37 & 1411.39 & 20.204 &    1.102 \\
 sc2 & 3191 &        & 18:04:26.22 & -28:46:49.4 &    -2.6 &    33.0 & 0.2390 &  19.76 &  583.10 & 19.427 &    0.899 \\
 sc2 & 4003 &        & 18:04:38.59 & -28:38:39.6 &     6.2 &   183.9 & 0.2990 &  16.51 &  931.29 & 17.301 &    0.687 \\
\enddata
\end{deluxetable}

\begin{deluxetable}{lrrccrrrrrrr}
\tablecaption{\label{tab:vis_lens}The same as
Table~\ref{tab:auto_lens} for events recovered in the visual search. 
}
\tablewidth{17.5cm}
\tablehead{
\colhead{Field} &
\multicolumn{2}{c}{Star} &
\colhead{~~~~~~$\alpha_{2000}$~~~~~} &
\colhead{$\delta_{2000}$} &
\colhead{~~~~~~~$F_0$} &
\colhead{$F_1$} &
\colhead{$u_0$} &
\colhead{$t_0$} &
\colhead{$t_{\rm max}$} &
\colhead{~~~~~~~$I_0$~~} &
\colhead{~~~$f$~~~} \\
& \colhead{DIA} & \colhead{OGLE} & & & & & & & & &
}
\startdata
 sc1 & 1827 &        & 18:02:07.16 & -30:02:11.7 &   -10.0 &   255.6 & 1.3514 &  40.69 & 1457.39 & 17.560 &    1.246 \\
 sc1 & 2096 &        & 18:02:24.89 & -29:58:44.3 &     0.2 &   701.2 & 1.6048 &   8.47 & 1080.94 & 16.293 &    1.064 \\
 sc1 & 2411 &        & 18:02:42.94 & -29:55:02.3 &    -1.4 &    17.9 & 0.1860 &   9.17 &  602.82 & 18.661 &    0.245 \\
 sc1 & 2624 &        & 18:02:45.02 & -29:52:59.9 &    -0.4 &    97.4 & 0.8416 &   2.89 &  576.66 & 19.388 &    2.651 \\
 sc1 & 2948 &        & 18:02:11.47 & -29:48:24.0 &    -5.7 &    53.2 & 0.4396 &  31.65 &  901.67 & 18.340 &    0.562 \\
 sc1 & 3087 &        & 18:02:38.16 & -29:47:44.2 &     0.3 &   466.6 & 1.2134 &   3.22 &  967.60 & 17.474 &    2.101 \\
 sc1 & 3327 &        & 18:02:47.14 & -29:44:34.5 &    -1.4 &    77.3 & 0.9806 &  29.72 &  602.27 & 18.129 &    0.642 \\
 sc1 & 3347 &        & 18:03:04.88 & -29:45:06.7 &    -1.2 &     6.5 & 0.0201 &  66.89 &  636.06 & 19.663 &    0.220 \\
\enddata
\end{deluxetable}

\begin{deluxetable}{lrrccrrrrrrr}
\tablecaption{\label{tab:other_lens}The same as Table~\ref{tab:auto_lens}
for additional microlensing events recovered through cross identification
with the OGLE catalog.}
\tablewidth{17.5cm}
\tablehead{
\colhead{Field} &
\multicolumn{2}{c}{Star} &
\colhead{~~~~~~$\alpha_{2000}$~~~~~} &
\colhead{$\delta_{2000}$} &
\colhead{~~~~~~~$F_0$} &
\colhead{$F_1$} &
\colhead{$u_0$} &
\colhead{$t_0$} &
\colhead{$t_{\rm max}$} &
\colhead{~~~~~~~$I_0$~~} &
\colhead{~~~$f$~~~} \\
& \colhead{DIA} & \colhead{OGLE} & & & & & & & & &
}
\startdata
 sc3 & 7783 & 590098 & 17:53:47.68 & -29:34:45.1 &    -2.8 & 13898.2 & 2.6588 &   2.17 &  579.35 & 17.293 &   50.614 \\
 sc5 & 2859 & 244353 & 17:50:36.09 & -30:01:46.6 &  -320.1 &   377.8 & 0.0894 & 248.33 &  529.46 & 18.351 &    3.934 \\
 sc5 & 4321 & 152125 & 17:50:13.29 & -29:49:56.9 &   -15.7 &  2084.6 & 0.0568 &  68.13 & 1580.15 & 15.075 &    0.998 \\
sc33 & 3764 & 164492 & 18:05:05.35 & -28:34:42.5 &  -229.7 &    31.2 & 0.0231 & 648.86 & 1365.66 & 18.093 &    0.241 \\
sc37 & 8176 & 321846 & 17:52:20.49 & -29:31:15.4 &    -1.2 &   702.4 & 1.7315 &  26.30 &  530.49 & 17.043 &    2.107 \\
sc40 & 2895 & 434222 & 17:51:10.76 & -33:03:44.1 &  -175.0 &    79.9 & 0.2204 & 435.88 & 1326.99 & 16.060 &    0.092 \\
sc44 & 1980 &  20145 & 17:48:56.53 & -30:11:40.8 &  -205.6 &  1219.0 & 0.7573 &  89.54 &  607.03 & 15.837 &    1.194 \\
sc44 & 2050 & 172247 & 17:49:33.12 & -30:11:56.1 &    -6.3 &   228.1 & 1.3794 &  48.98 & 1360.95 & 17.874 &    1.536 \\

\enddata
\end{deluxetable}

}

\end{document}